\pdfoutput=1 
\documentclass[12pt]{article}
\pdfoutput=1
\usepackage{latexsym}
\usepackage{graphicx}
\usepackage{amsmath}
\usepackage{amsfonts}
\usepackage{amssymb}
\usepackage{color}
\usepackage{bm}
\numberwithin{equation}{section}

\newcommand{\adda}{Raman Research Institute, Bangalore 560080, India.}
\newcommand{\addb}{Graduate School of Science, University of Tokyo, 113-0033, Japan}
\newcommand{\addd}{Institute of Physics, University of Augsburg,
 Universit\"atsstrasse 1, D-86135 Augsburg, Germany}
\newcommand{\adde}{Department of Physics and Centre for Computational
Science and Engineering, National University of Singapore, Singapore
117546 }
\newcommand{\addh} {Max Planck Institue for the Physics of Complex systems,
N\"othnitzer Strasse 38, D-01187 Dresden, Germany}

\newcommand{\beq }{\begin{eqnarray}}
\newcommand{\eeq }{\end{eqnarray}}
\newcommand\dg{\dagger}
\newcommand\al{\alpha}
\newcommand\nn\nonumber
\newcommand\f\frac
\newcommand\bG{{\bm G}}

\newcommand\bV{{\bm V}}
\newcommand\bM{{\bm M}}
\newcommand\bK{{\bm K}}

\newcommand\bGam{{\bm \Gamma}}
\newcommand\bSig{{\bm \Sigma}}
\newcommand\bOm{{\bm \Omega}}
\newcommand\blam{{\bm \lambda}}
\newcommand\cH{{\cal{H}}}
\begin{document}

\title{Nonequilibrium  density matrix description of steady state quantum transport}
\author{Abhishek Dhar \thanks{\adda}, Keiji Saito \thanks{\addb} and
Peter H\"anggi \thanks{\addd} \thanks{\adde} \thanks{\addh} }
\date{\today}
\maketitle
\begin{abstract}
With this work we investigate  the stationary {\it nonequilibrium} density matrix of current carrying nonequilibrium steady states
of in-between quantum systems that are connected to  reservoirs.
We describe the analytical procedure to obtain the explicit result for the reduced
density matrix of quantum transport  when the system,
the connecting reservoirs and, as well,  the system-reservoir interactions are  described by  quadratic Hamiltonians.
Our  procedure is detailed for both,   electronic transport described by
the tight-binding Hamiltonian and for phonon transport described by harmonic
Hamiltonians. For  the special case of weak system-reservoir couplings, 
a more detailed description of the steady-state density matrix is obtained. 
Several paradigm transport setups for inter-electrode electron transport and low-dimensional phonon heat flux are elucidated.
\end{abstract}
\vskip .5 true cm

\medskip
\noindent

\newpage

\section{Introduction}
\label{intro}

The theory of equilibrium statistical mechanics, as pioneered by
Boltzmann and Gibbs,  provides the prescription for the appropriate
density matrix (or density operator)-description  of a system that is kept under various external
constraints. Thus, for
systems kept in isolation the microcanonical
distribution yields the appropriate density matrix, while for systems in weak
contact with a thermal and particle reservoir the grand-canonical
density matrix describes the statistical state of the system.
For classical systems, equilibrium statistical physics is governed  by the
phase space distribution of the system. A knowledge of the density
matrix or the phase space distribution then enables one to find various
equilibrium and also close to equilibrium properties  of a system, as exemplified, for example,  via linear
response theory.

For systems taken far away from equilibrium there exists no general procedure in  obtaining
its density matrix. Particularly, this  holds true  for systems that have reached steady
states. For classical Hamiltonian systems described by a Markovian stochastic
dynamics the steady state is determined by the stationary solution of the corresponding
master equation; e.g. it is given by the stationary probability density of the Fokker-Planck generator for continuous
Markovian processes \cite{PR}. Apart from specific situations, however, for example (i) in  the presence of symmetries such as (strict) detailed balance, or (ii) a single variable state space dynamics \cite{PR}, the task of finding the closed form solution of such master equations presents a profound challenge which typically can be obtained only by the usage of extensive numerical simulations or algorithms.

In this context we remind the readers that even for the case of a system being in contact with a {\it single} bath the corresponding canonical equilibrium is typically not of the common Boltzmann-Gibbs structure, as encoded with the exponential of the (negative) bare system Hamiltonian and inverse temperature. The latter structure holds rigorously true for weak coupling. In presence of strong coupling, however, the corresponding thermal (generalized canonical) density operator then typically involves a temperature-dependent "Hamiltonian of mean force" \cite{CampisiPRL} which includes entropic contributions that  explicitly depend on the system-bath coupling strength.

Regrettably, no such general concept as the canonical Boltzmann-Gibbs density matrix structure in terms of the  bare (or even modified Hamiltonian of mean force) is available when the open system is subjected to  steady state transport. Put differently, there are no generic results known for  stationary nonequilibrium statistics. This latter situation in fact is not only substantially  more complex but presently is also less researched. It is thus of outmost importance to gain  further insight into this objective of obtaining the  underlying nonequilibrium density matrices that govern quantum and/or  classical transport. For example, the explicit form of a corresponding nonequilibrium density matrix not only determines the linear response due to an additional external perturbation of such a nonequilibrium steady state (NESS), but also its higher order response functions.

A particular, exactly solvable case is that of heat conduction occurring in a  one-dimensional
ordered harmonic chain when  connected to two baths at different
temperatures. If the two baths are modeled therein as being {\it stochastic}
with corresponding stochastic forces acting on the system of interest, the  exact nonequilibrium steady state phase space
distribution for this problem was evaluated by Rieder, Lebowitz and Lieb
\cite{RLL67}. An extension to the case of higher dimensions was later obtained
by Nakazawa \cite{Nakazawa}.  Heat conduction in quantum harmonic oscillator
chains has been studied by several authors
\cite{zurcher90,saito00,dharshastry03,segal03}, but thus far no explicit results are known
for  the precise form of the  quantum mechanical {\it steady state} density
matrix. Some formal results for the NESS density matrix  of general
quantum mechanical systems have been obtained in the
works of Zubarev \cite{zubarev74} and McLennan \cite{mclennan63} and have more
recently been discussed in specific models \cite{pillet,tasaki,platini}.

The problem of obtaining the NESS in explicit form presents a formidable challenge already for classical open systems (see above discussion).   This objective therefore is typically even  more intricate for a quantum NESS. Indeed, in presence of general nonlinear interactions this task is simply inaccessible without invoking also  extensive and cumbersome numerical means and methods. To obtain  general analytic insight
over  whole parameter ranges thus necessitate to confine the  objective to stylized situations only, that allow for  explicit closed
form calculations.

With this work, we consider generic setups for steady state quantum transport described by a bilinear Hamiltonian. The aim is to  find systematic procedure for obtaining explicit results for the  NESS density matrix for
this class of systems. We demonstrate that it is
possible to obtain the complete NESS density matrix explicitly.
We also show that when the coupling strength between the system and reservoirs are
extremely weak, the NESS density matrix is given by an effective Gibbs state where each mode
is formally only in equilibrum with a mode-dependent effective temperature which depends, however,
in a complex manner on  both bath temperatures.

We consider two generic setups for stationary nonequilibrium quantum transport, a first one involving
fermions and the other one  bosons as carriers.
The first setup consists of electron and heat transport in a
fermion setup of non-interacting particles  which are connected to fermionic baths at
different temperatures and chemical potentials. The second scenario consists of
heat conduction occurring in harmonic crystals connected to oscillator baths kept at
different temperatures. For both these problems it is known from prior
studies, using various approaches such as the nonequilibrium Green's function
formalism \cite{jauho94,haugjauho}, the
quantum Langevin equations approach \cite{dharsen06,DharRoy06,Dhar08}, and the
$C^\star$-algebra approach \cite{tasaki01}, that it
is  possible to express all two point correlations in the NESS in terms
of appropriate Green's functions. Because these systems are non-interacting it
is evident that the two-point correlations contain necessary information on all
higher-point correlations and hence should completely specify the NESS. With this study
we give the procedure for finding the explicit NESS density matrices from a
knowledge of the two-point correlations in these systems.

For the  case of a weak-coupling among system and the baths we are
able to obtain explicit results. We further present explicit examples in simple one-dimensional models
which illustrate our general procedure and also demonstrate the
accuracy of the weak-coupling approximation.

The outline of the paper is as follows. In Sec.~(\ref{sec:genSS}) we present the
general procedure for construction of the NESS for the electron and
phonon transport problems. The special case of weak coupling between system-bath  is discussed
in Sec.~(\ref{sec:wc}). In Sec.~(\ref{sec:examples}) we discuss some
illustrative  examples of models where both system and reservoirs are
taken to be one-dimensional chains.  Finally we end with a discussion
in Sec.~(\ref{sec:discussions}).

\section{Construction of Steady State Density Matrix}
\label{sec:genSS}
In this section, we outline the general procedure to obtain the steady
state density matrix  in quantum transport described by a bilinear Hamiltonian.
We focus on electric
conduction as an example of fermionic transport, and phononic
heat conduction as bosonic transport.
Because the resulting NESS density matrix becomes be Gaussian, the pertinent procedure  fist is in finding the explicit
form of the two-point correlation functions of physical quantities and next relating these to the Gaussian distribution.

\subsection{Steady state density matrix for non-interacting electron transport}
\label{sec:ssel}
We consider the typical setup of transport in the Landauer approach
wherein a system is connected to two reservoirs initially kept at different
temperatures and chemical potentials. At long times the system reaches a
nonequilibrium steady state with a mean rate of flow of charge and energy
current. One starts out by writing the full Hamiltonian of the system plus
reservoirs and here we consider a tight-binding approach of
non-interacting electrons. We use the following notation:
for sites on the system ($S$) we shall use the integer indices $l,m,n, \cdots$; for sites on the left reservoir ($L$)
we employ the Greek indices $\alpha, \nu$; and finally, for sites on
the right reservoir ($R$) we use the primed Greek indices $\alpha', \nu'$.
We consider quantum transport with the following overall Hamiltonian reading:
\beq
{\mathcal{H}} &=&  {\cal{H}}_S ~+~ {\cal{H}}_L ~+~ {\cal{H}}_R ~+~ {\mathcal{H}}_{LS} ~+~{\mathcal{H}}_{RS} ~,
\label{ham} \\
{\rm where} \quad {\cal{H}}_S &=& \sum_{lm} ~{\bm H}_{lm} ~c_l^\dg c_m,~~{\cal{H}}_L ~=~
\sum_{\al \nu} ~{\bm H}^L_{\al \nu} ~c_\al^\dg c_\nu,~~ {\cal{H}}_R ~=~ \sum_{\al' \nu'} ~
{\bm H}^R_{\al' \nu'} ~c_{\al'}^\dg c_{\nu'} ~, \nn \\
{\mathcal{H}}_{LS} &=& \sum_{l \al} ~ {\bm H}^{LS}_{l \al} ~[~c_l^\dg c_\al ~+~
 ~c_\al^\dg c_l ~] ~, \nn \\
{\mathcal{H}}_{RS} &=& \sum_{l \al'} ~ {\bm H}^{RS}_{l \al'} ~[~c_l^\dg c_{\al'} ~+~
~c_{\al'}^\dg c_l ~] ~, \nn
\eeq
where $c^\dg,~c$ denote creation and annihilation operators satisfying the usual
fermionic anti-commutation rules and we  assume that the matrices ${\bm H},
{\bm H}^L, {\bm H}^R$ are symmetric and real-valued while ${\bm H}^{LS}, {\bm H}^{RS}$ are  real-valued. In the above setup we assume
that the system possesses a
finite number of lattice sites $N$ while the left and right reservoirs have
$N_L$ and $N_R$ sites which will eventually be made infinite.  The
parts $\cal{H}_S$, ${\cal H}_L$ and ${\cal H}_R$ denote the Hamiltonians of the
isolated system, left and right reservoirs respectively, while $\mathcal{H}_L$
and $\mathcal{H}_R$ describe the coupling of the left and right reservoirs
to the system, which have been taken to be real.
To obtain a NESS for the system we consider  an initial state at time
$t=t_0$ given by the following product density matrix:
\beq
\rho(t_0)= \rho^0_S \otimes \rho^0_L \otimes \rho^0_R~, \label{idm}
\eeq
where $\rho^0_L \sim e^{-({\mathcal{H}}_L-\mu_L \mathcal{N}_L)/T_L}$
($\rho^0_R \sim e^{-({\mathcal{H}}_R-\mu_R \mathcal{N}_R)/T_R}$) is the equilibrium grand-canonical density matrix of
the left (right) reservoir,  corresponding to temperature $T_L$ ($T_R$) and
chemical potential $\mu_L$ ($\mu_R$) with $\mathcal{N}_{L},\mathcal{N}_R$ the total number
operators, while $\rho^0_S$ denotes an arbitrary initial
density matrix for the system.
We then time-evolve the whole system with the full Hamiltonian given in
Eq.~(\ref{ham}) so that at time $t$ the full density matrix is given by:
\beq
\rho(t)= e^{i \mathcal{H} (t-t_0)/\hbar} \rho(t_0) e^{- i \mathcal{H} (t-t_0)/\hbar }~.
\eeq
Our principal objective is the long time limit of the steady state reduced density matrix for the system under consideration, i.e.,
\beq
\rho_S= \lim_{t_0 \to - \infty}\; \lim_{\text{baths} \to \infty} \;{\rm Tr}_{L,R}~ \rho(t)~,
\eeq
where the trace, ${\rm Tr}_{L,R}$, is over all the degrees of freedom of the two baths.
In doing so we implicitly assume that the quantum transport setup is so that (i) it possesses  a long time limit in the limit of infinite many bath degrees of freedom and (ii) that the interactions within the quantum system and the interaction with the bath degrees of freedom are such that the emerging asymptotic nonequilibrium steady state density matrix  indeed is time-independent.

Let us introduce the two-point correlation function
\beq
\langle\, c_{m}^{\dagger} c_l \,\rangle= {\rm Tr}_S [ c_m^\dg c_l \, \rho_S ]= {\rm Tr} [
  c_m^\dg  c_l \, \rho ]
\eeq
where the first trace, ${\rm Tr}_S$, is over system degrees of freedom and the
second trace is over all degrees of freedom.
Because of the quadratic form of the total Hamiltonian the two-point
correlations in the NESS  can be exactly calculated using  various methods
\cite{caroli,meir,dharsen06}.

The correlations can be expressed in terms of the following Green function:
\beq
{\bm G}^\pm(\omega)=\f{1}{\hbar\omega -{\bm H}-{\bm \Sigma}_L^\pm(\omega)-{\bm \Sigma}_R^\pm(\omega)} ~,
\label{self}
\eeq
where $\Sigma^\pm_{L}$ and $\Sigma^\pm_R$ are self-energy terms which
model the effect of the infinite reservoirs on the isolated
system Hamiltonian.
The self energies can be written in terms of the isolated
reservoir Green functions ${\bm g}^\pm_{L}(\omega)=[\hbar\omega \pm i
  \epsilon-{\bm H}^{L}]^{-1}$, ${\bm g}^\pm_{R}(\omega)=[\hbar\omega \pm i
  \epsilon-{\bm H}^{R}]^{-1}$
and the coupling matrices ${\bm H}^{LS}$ and ${\bm H}^{RS}$, reading
\beq
{\bm \Sigma}^\pm_L(\omega)={\bm H}^{LS} {\bm g}^\pm_L(\omega) {{\bm
    H}^{LS}}^{\dg},~~~~{\bm \Sigma}^\pm_R(\omega)={\bm H}^{RS} {\bm g}^\pm_R(\omega)
{{\bm H}^{RS}}^{\dg} ~.
\eeq
Let us next define ${\bm \Gamma}_L(\omega)=[{\bm \Sigma}^-_L-{\bm \Sigma}^+_L~]/(2i)~,~{\bm
  \Gamma}_R(\omega)= [{\bm \Sigma}^-_R- {\bm \Sigma}^+_R]/(2i)$.
With these definitions one finds the following expressions for the
steady state correlation matrix:
\beq
{\bm C}_{ml} &=& \langle~ c_m^\dg ~ c_l~ \rangle \nn \\
&=& \int_{-\infty}^\infty d\omega ~\frac{\hbar}{\pi}~[~({\bm G}^+ {\bm \Gamma}_L
  {\bm G}^-)_{lm}~
f(\omega,\mu_L,T_L) ~+~ ({\bm G}^+{\bm \Gamma}_R {\bm G}^-)_{lm} ~f(\omega,\mu_R,T_R)~] ~,~~~~\label{negfres}
\eeq
where $f(\omega,\mu,T)=1/[e^{\beta(\hbar \omega-\mu)}+1]$ denotes the Fermi function.

We demonstrate next how the
NESS density matrix $\rho_S$ can be fully expressed in terms of these
correlations.
Note that the matrix ${\bm C}$ is Hermitian, since at any time ${\rm Tr}[~c_l^\dg~
  c_m ~\rho(t)] = {\rm Tr}[~c_m^\dg~   c_l~ \rho(t)~]^* $, where (*) indicates complex conjugation. This result can also be
directly verified from the form in Eq.~(\ref{negfres}).
Consequently the matrix ${\bm C}$ can be diagonalized with a unitary
matrix ${\bm U}$ to read:
\beq
{\bm U}^{\dagger} {\bm C} {\bm U} &=&{\bm D}~ =~
 {\rm Diag} (d_1 , d_2 , \cdots , d_{N-1}, d_{N}) , \label{unitary}
\eeq
where the matrix ${\bm D}$ is the diagonal matrix.
Using the unitary transformation, we define  new fermionic operators as
\beq
c_s ' =  \sum_{l} {\bm U}_{l,s}~c_l ,~~~s=1,\ldots,N~.
\eeq
Obviously, these new fermionic operators preserve the anti-commutation relations,
$\{ c_s' , c_{s'}'^{\dagger} \}=\delta_{s,s'}$.
The steady state density matrix is a diagonal matrix in terms of these new fermion operators.
Note that the two-point correlation of new fermionic operators
read $\langle c_{s}'^{\dagger} c_{s'}'\rangle = \delta_{s,s'} d_s$. From this we  find the corresponding effective
Fermi-Dirac distribution for each fermion $s$. Consequently,
the steady state matrix $\rho_{S}$ is formally given by
\beq
\rho_{S} &=& \prod_{s=1}^N
{ \exp \left[ -a_s c_{s}'^{\dagger} c_{s} ' \right]
\over [1 + \exp ( -a_s )]}  \label{fermion_ss1}\\
&=& \frac{\exp \Bigl[ -\sum_{l,m} c_l^{\dagger} {\bm A}_{l,m} c_m  \Bigr]
}{ \prod_{s=1}^N \left[ 1 + \exp(-a_s) \right]} \, , \label{fermion_ss2}
\\
{\bm A} &=&  {\bm U}^{\ast} \, {\rm Diag}(a_1,a_2, \cdots, a_{N-1},a_N )
\, {\bm U}^{T}  \, , \label{fermion_ss3}\\
a_s &=& \ln \left( d_s^{-1} -1 \right) . \label{fermion_ss4}
\eeq
To obtain Eq. (\ref{fermion_ss4}) we used   the relation $\langle~
c_s'^{\dagger} ~c_s '~ \rangle = d_s = 1/[\exp (a_s) + 1]$.  This completes
our derivation of the expression for the steady state density matrix
for noninteracting electron transport.

\subsection{Steady state density matrix  for noninteracting phonon transport}
\label{sec:ssph}
We next consider heat conduction in general harmonic networks.  Examples of
such a system are  dielectric crystals for which the harmonic crystal
provides a very good description.  As before we again consider the usual Landauer-like
framework of a system connected to two reservoirs kept at different
temperatures \cite{dharshastry03,segal03}. The reservoirs are themselves modeled as collections of harmonic
oscillators.  Let us assume that the system has $N$ Cartesian positional
degrees of freedom $\{x_l\}$, $l=1,2\ldots,N$ with corresponding momenta
$\{p_l\}$. These satisfy the usual commutation relations $[x_l,p_m]=i \hbar
\delta_{l,m}$ and $[x_l,x_m]=[p_l,p_m]=0$. Similarly the left reservoir degrees of freedom are denoted by
$\{x^L_\alpha,p^L_\alpha\}$,~ $\alpha=1,\ldots,N_L$ and the right reservoirs by
$\{x^R_{\alpha'}, p^R_{\alpha'}\}$, ~$\alpha'=1,\ldots,N_R$. We will use the
vector notation $X^T=(x_1,x_2,\ldots,x_N)$,~
$P^T=(p_1,p_2,\ldots,p_N)$ and similarly $X^L,X^R, P^L,P^R$. The
 general  Hamiltonian for the system coupled to harmonic reservoirs is then given by:
\beq
{\mathcal{H}} &=&  {\cal{H}}_S ~+~ {\cal{H}}_L ~+~ {\cal{H}}_R ~+~ {\mathcal{H}}_{LS} ~+~{\mathcal{H}}_{RS} ~,
\label{ham2} \\
{\rm where}~~\mathcal{H}_S &=& \f{1}{2} P^T~{\bm M}^{-1} ~P +\f{1}{2}
X^T ~{\bm K}~ X~, \nn \\
\mathcal{H}_L &=& \f{1}{2}[{P^L}]^T~[{\bm M}^L]^{-1} ~P^L +\f{1}{2}
[{X^L}]^T ~{\bm K}^L X^L~, \nn \\
\mathcal{H}_R &=& \f{1}{2}[{P^R}]^T~[{\bm M}^R]^{-1} ~P^R +\f{1}{2}
[{X^R}]^T ~{\bm K}^R X^R~, \nn \\
{\mathcal H}_{LS} &=&  X^T~{\bm K}^{LS}~X^L~,~~ {\mathcal H}_{RS}=  X^T~{\bm  K}^{RS}~X^R~, \nn
\eeq
where ${\bm M},~{\bm M}^L,~{\bm M}^R$ and ${\bm K},~{\bm
  K}^L,~{\bm K}^R$ denote respectively the mass matrix and the
force-constant matrix of the system, left reservoir and right
reservoir, while ${\bm K}^{LS}$ and ${\bm K}^{RS}$ denote the linear
coupling coefficients between the two reservoirs and the system.

Again we consider the time  evolution of the coupled system plus reservoirs
starting from an initial product density matrix of the form
Eq.~(\ref{idm}) with $\rho^0_L \sim \exp (-{\cal{H}}_L/k_B T_L)$ and
$\rho^0_R \sim \exp (-{\cal{H}}_R/k_B T_R)$ and the system being in an
arbitrary initial state. At long times the system reaches a NESS
described the reduced density matrix $\rho_S={\rm Tr}_{L,R}
\rho(t\rightarrow \infty)$.
In order to construct $\rho_S$ , we start with
defining the appropriate correlation matrix as in the previous section for electron transport.
In doing so we consider  the $2N\times 2N$ covariance matrix defined with the column vector
 ${ \varphi} = (x_1, \cdots , x_N , p_1, \cdots , p_N )^T $:
\beq
{\bm C} =
\Bigl\langle
{ \varphi} { \varphi}^{T}
\Bigr\rangle = {\rm Tr}_S [~\varphi \varphi^T ~\rho_S~]~.
\eeq
For this covariance matrix, we write the symmetric and anti-symmetric parts as
\beq
{\bm C}_S &=& {1\over 2} \left( {\bm C} +{\bm C}^{T} \right) \, , \\
{\bm C}_A &=& {1\over 2} \left( {\bm C} -{\bm C}^{T} \right) = {i\hbar \over 2} {\bm J} \\
{\bm J} &=& \left(
\begin{array}{rr}
{\bm 0} & ~{\bm 1} \\
-{\bm 1} & ~{\bm 0}  \\
\end{array}
\right),
\eeq
where ${\bm 1}$ and ${\bm 0}$ are respectively $N\times N$ identity and zero matrices.
The matrix expression of anti-symmetric part ${\bm C}_A$ is automatically determined by
commutation relations between coordinate and momentum variables.

The symmetric part of covariance matrix
${\bm C}_S$ is given by
\beq
{\bm C}_S &=&
\left(
\begin{array}{cc}
\langle { X}{X}^T \rangle  &
{1\over 2}{\langle { X} { P}^T  + [{ P} { X}^T]^T \rangle } \\
{1\over 2}{\langle { X} { P}^T  + [{ P} { X}^T]^T \rangle }
&
\langle { P} { P}^T \rangle
\end{array}
\right) ~.
\eeq
As for the electron case these correlations are known in terms of the
following phonon Green function:
\beq
{\bm G}^{\pm} &=& \frac{1}{-{\bm M} \omega^2 + {\bm K} - {\bm \Sigma}_L^{\pm} - {\bm \Sigma}_R^\pm } \, ,
\eeq
where the self-energies can be expressed  in terms of the isolated reservoir Green
functions ${\bm g}^\pm_{L}(\omega)= [~-{\bm M}^{L}(\omega\pm i \epsilon)^2+{\bm
    K}^{L}~]^{-1}  $ , ${\bm g}^\pm_{R}(\omega)= [~-{\bm M}^{R}(\omega \pm i
  \epsilon)^2+{\bm     K}^{R}~]^{-1}  $
 and the coupling elements ${\bm K}^{LS},~{\bm K}^{RS} $. These self energies thus read

\beq
{\bm \Sigma}^\pm_L(\omega)={\bm K}^{LS}~ {\bm g}^\pm_L(\omega)~ [{\bm
    K}^{LS}]^{T},~~~~{\bm \Sigma}^\pm_R(\omega)={\bm K}^{RS}~ {\bm g}^\pm_R(\omega)~
[{\bm K}^{RS}]^{T} ~.
\eeq
Defining ${\bm \Gamma}_L(\omega)={\rm Im} [~{\bm \Sigma}^+_L~]~,~{\bm
  \Gamma}_R(\omega)= {\rm Im} [~ {\bm \Sigma}^+_R~]$, we find
\cite{DharRoy06,wang06,yamamoto06}:
\beq
\langle {X}{X}^T \rangle &=&
\int_{-\infty}^{\infty} d \omega {\hbar \over 2\pi } \sum_{a=L,R}
 {\bm G}^+ {\bm \Gamma}_a {\bm G}^- ~g(\omega,T_a) ~, \nn \\
\langle { P}{ P}^T \rangle &=&
\int_{-\infty}^{\infty} d \omega {\hbar \omega^2 \over 2\pi } ~\sum_{a=L,R}
{\bm M}{\bm G}^+ {\bm \Gamma}_{a} {\bm G}^- {\bm M}~ g(\omega,T_a) ~ , \nn \\
{1\over 2}{\langle { X} { P}^T  + [{ P} { X}^T]^T \rangle }
&=& \int_{-\infty}^{\infty} d \omega
{i\hbar\omega \over \pi } \sum_{a=L,R}
 {\bm G}^+ {\bm \Gamma}_{a} {\bm G}^-{\bm M}  ~ g(\omega,T_a) ~,~~~~  \label{harmCM}
\eeq
where $g(\omega,T)=\coth (\hbar\omega /2 k_B T)$.

We next show how the steady state density matrix can be expressed in terms of
the correlation matrix. For this it is necessary to consider symplectic transformations \cite{gosson}. We first introduce the symplectic matrix ${\bm S}$, satisfying
\beq
{\bm S} {\bm J} {\bm S}^{T} &=& {\bm J} \, , \label{symplectic1}\\
{\bm S} {\bm C}_S {\bm S}^{T} &=& {\bm D}~ =~ {\rm Diag}~(d_1, \cdots, d_N , d_1, \cdots , d_N )  .
\label{symplectic2}
\eeq
The procedure to find ${\bm S}$ is detailed  in  the Appendix~(\ref{appa}).

By using the  symplectic transformation with the matrix ${\bm S}$, the new
operators   ${ \varphi}' = (x_1', \cdots , x_N' , p_1', \cdots , p_N' )^T $
are defined as:
\beq
{\varphi}'_s &=& \sum_{l=1}^N {\bm S}_{s,l}~ {\varphi}_l~ ~~~~,s=1,\ldots,N~.
\eeq
The most important property of the
the symplectic transformation, following from Eq.(\ref{symplectic1}),  is that
it preserves the commutation relations and we have $[x_s,p_{s'}]=i \hbar \delta_{s,s'}$ and
$[x_s,x_{s'}]=[p_s,p_{s'}]=0$.
The steady state density matrix can then be written in terms of these
new operators and we end up with the general main result:
\beq
\rho_{S} &=&
\prod_{s=1}^{N}
\frac{ \exp \left[ -a_s (x_s'{} ^2 + p_s'{}^2 ) \right]}
{{ Z}_s}  \label{boson_ss1} \\
&=&  \frac{\exp \left[ -{\varphi }^T {\bm  A} { \varphi} \right]}
{\prod_{s=1}^{N} {Z}_s}    \, , \label{boson_ss2a}\\
{\bm A} &=&  {\bm S}^{T}~ {\rm Diag} ~(a_1 , \cdots , a_N , a_1, \cdots , a_N ) ~{\bm S} \, , \label{boson_ss2b}\\
{ Z}_s &=& \left[~ 2 \sinh (\hbar a_s )  ~\right]^{-1}\, ,\label{boson_ss3}\\
a_s &=& \hbar^{-1} \coth^{-1} (2 d_s / \hbar )~ \label{boson_ss4}.
\eeq
In computing the normalization factor ${ Z}_s$, we have used the second
quantization representation
$x_{s}' =  \sqrt{\hbar \over 2} (b_s^{\dagger}  + b_s ) ~,
p_{s}' = i \sqrt{\hbar \over 2} (b_s^{\dagger} - b_s ) $,
where $b_s$ and $b_s^{\dagger}$ satisfy $[ b_s , b_{s'}^{\dagger} ]
=\delta_{s,s'} $.
Then we obtain the expression $a_s (x_s'{}^2 + p_s'{}^2) = 2\hbar a_s
(b_s^{\dagger} b_s + 1/2)$.  The relation between $d_s$ and $a_s$ is then
found by looking at the averages $\langle~ x_s'{}^2~ \rangle$ and $\langle~
p_s'{}^2 ~\rangle$:
\beq
\langle ~x_s'{}^2~ \rangle =
\langle~ p_s'{}^2~ \rangle = {\hbar\over 2}\coth(\hbar a_s ) = d_s .
\eeq

Finally we also consider here the classical limit $\hbar\to 0$. In this limit, we have
the simple  relation $d_s = 1/(2 a_s)$. Then, the matrix ${\bm A}$
is given by
\beq
{\bm A} &=& {1\over 2} {\bm S}^{T} {\bm D}^{-1} {\bm S} ~~~~~~~(\hbar\to 0).
\eeq
Hence, from the relation
$ {\bm D}^{-1}  =
({\bm S} {\bm C} {\bm S}^{T} )^{-1} = ({\bm S}^T)^{-1} {\bm C}^{-1} {\bm
  S}^{-1} $, we find the following expression of the matrix ${\bm A}$ in the
classical limit:
\beq
{\bm A} &=& {1\over 2} {\bm C}^{-1}  ~~~~~~~(\hbar\to 0).
\eeq
Thus we recover the form that is expected for a general  Gaussian
probability measure. We note that in the classical case, for Gaussian white
noise  reservoirs, the correlation matrix ${\bm C}$ can be explicitly determined for
ordered  harmonic lattices \cite{RLL67,Nakazawa}. For arbitrary harmonic
networks, they are given by the high temperature limit of Eqs.(\ref{harmCM}),
with appropriate choices of the bath spectral functions.
Finding the inverse of the correlation matrix presents, however, a more difficult
task.

\section{Weak coupling limit}
\label{sec:wc}
In this section, we consider the special case of  a weak coupling
between the system and reservoirs. We note that it is essential that the
weak-coupling limit is taken {\it after} the coupled system-reservoirs have evolved
for an infinte time and thus reached the NESS.
Generally, when the coupling strength is  weak, the density matrix can be
expanded in terms of the coupling strength. In this case, the zeroth order
term in the coupling strength determines the overall structure of the electron
density profile  in the electron conduction case, and the
temperature profile in the case of phonon heat conduction. The higher order
terms of the expansion  determine the amount of current flowing in the system.
Therefore, although the coupling strengths must be finite for finite current, even
the  zeroth order  contribution in the expansion of the density matrix carries
important information on the steady state.
In this section, we focus on the $0$-th order  contribution in the weak coupling
expansion of the steady state density matrix, which we here refer to as the
density matrix in the {\em weak coupling limit}.
We emphasize that at no instant we switch off the coupling strength which is
always kept finite, but small.

On decreasing the coupling strength, the current decreases; however,   even in the
limit of zero current, the
{\em steady state density matrix is non-trivial and different
from the equilibrium density matrix}.  In fact, we will find that the
NESS is non-unique in the sense that it depends on the
way the system-coupling strengths are made to vanish.
For the case where the temperatures and chemical potentials of the two
reservoirs are chosen equal one obtains, in  the weak-coupling limit, the expected
equilibrium grand-canonical (for electron  case) and canonical (for phonon
case) distributions.

\subsection{Electron transport}
\label{wc-electron}

We first note that the system's Hermitian Hamiltonian matrix ${\bm H}$
has the eigenvalue-equation $\sum_m {\bm H}_{l,m} {\bm V}_{m}(s) = {\bm \lambda}_{s} {\bm V}_l (s)$, hence
can be diagonalized by the unitary transformation $\bV$ as
\beq
\bV^\dg {\bm H} \bV = {\bm \lambda}~,~~~~~\bV^\dg \bV = {\bm I}~.
\eeq
Next we use the spectral decomposition:
\beq
\bG^+ &=& \bV  \bV^{-1} [\hbar\omega -{\bm H} -\bSig^+_L-\bSig^+_R]^{-1}
~[\bV^\dg]^{-1} \bV^\dg \nn \\
&=& \bV [\hbar\omega -\blam- \bV^\dg (\bSig^+_L + \bSig^+_R) \bV]^{-1}~\bV^\dg~. \label{gplus}
\eeq
From this it follows that in the weak coupling limit $\bSig_L^+,\bSig_R^+ \to
0$, the matrix element $\bG^+_{l,m}$ is effectively given by
\beq
\bG^+_{l,m}= \sum_{s} \f{\bV_l(s) \bV^*_m(s)}{\hbar\omega +\blam_s -i \langle~
  s~|~\bGam~|~s~\rangle  }~,
\eeq
where $\langle  ~ s~|~\bGam~|~s'~\rangle= \sum_{l,m} \bV^*_l(s) (\bGam)_{l,m}
\bV_m(s')$
and  $\bGam=\bGam_L+\bGam_R$.
It can be shown that the off-diagonal terms
of the inverse matrix in Eq. (\ref{gplus}) are of the order of the coupling strength. This contribution disappears, however,
in the following calculation of the correlation function, given this  weak coupling limit.
The real part of $\bSig^+_{L,R}$ is negligible compared to the remaining real parts and thus can be
dropped.
Hence we obtain:
\beq
\langle~ c^\dg_m~ c_l ~\rangle &=& \int_{-\infty}^\infty d\omega \f{\hbar }{ \pi} \sum_{a=L,R}
\sum_{j,k}
\bG^+_{l,k} (\bGam_a)_{k,j} \bG^-_{j,m}~f(\omega,\mu_a,T_a) \nn \\
&=& \int_{-\infty}^\infty d \omega \f{\hbar }{\pi} \sum_{a=L,R}
\sum_{s,s',j,k} \f{\bV_l(s) \bV^*_k(s)}{\hbar\omega -\blam_s -i \langle~
  s~|~\bGam~|~s~\rangle }  [ \bGam_a ]_{k,j} \nonumber \\
&& ~~~~~~~~~~~~~~~~~~~~~~~\times
\f{\bV_j(s') \bV^*_m(s')}{\hbar\omega - \blam_{s'} +i \langle
 ~ s'~|~\bGam~|~s'~\rangle  }~f(\omega,\mu_a,T_a)~. \nn
 \eeq
A careful examination of the limit $\langle~ s~|~\bGam_a~|~s~\rangle \to 0$
exhibits that only the terms $s=s'$ survive in the above summation, yielding:
\beq
\langle~ c^\dg_m~ c_l~ \rangle =  \int_{-\infty}^\infty d \omega \f{\hbar}{\pi} \sum_{a=L,R}
\sum_{s} \f{\bV_l(s)~ \langle~ s~|~ \bGam_a(\omega)~ |~s~ \rangle~ \bV^*_m(s) }{(\hbar\omega
  -\blam_s)^2 + \langle   ~s~|~\bGam(\omega)~|~s~\rangle^2 }
~f(\omega,\mu_a,T_a)~. \nn
\eeq
Next,  making use of the  identity
\beq
\lim_{\epsilon \to 0} \f{\epsilon}{(x-a)^2+\epsilon^2} = \pi  \delta(x-a)~, \nn
\eeq
we find:
\begin{eqnarray}
\langle~ c^\dg_m~ c_l ~\rangle  &=& \sum_s   \bV_l(s) \bV^*_m(s)~e_s \nn \\
{\rm where}~e_s &=& \sum_{a=L,R} \f{\langle~   s~|~\bGam_a~|~s~\rangle }{\langle
  ~s~|~\bGam~|~s~\rangle }~  f(\blam_s/\hbar,\mu_a,T_a)~ \label{wcelectron} \nn \\
&=& \gamma_L ~f(\blam_s/\hbar,\mu_L,T_L)+ \gamma_R ~f(\blam_s/\hbar,\mu_R,T_R)  ~, \nn \\ 
{\rm where}~~ \gamma_{L} &=&  \f{\langle   ~s~|~\bGam_{L}~|~s~\rangle }{\langle   ~s~|~\bGam~|~s~\rangle }~,~~~~\gamma_{R} =  \f{\langle   ~s~|~\bGam_{R}~|~s~\rangle }{\langle   ~s~|~\bGam~|~s~\rangle }=1-\gamma_L~. \nn
\end{eqnarray}
Note that, in the above expression,  the limit $\langle~   s~|~\bGam_a~|~s~\rangle \to 0 $ is always implied
and it is then evident that the ratios  $\gamma_L, \gamma_R$ depend on the way the couplings $\to 0$. From the form above we can interpret $e_s$ as an effective occupation probability of the  energy-level $\blam_s$ of the isolated system and this probability depends on the temperatures and chemical potentials of the two reservoirs.   
Defining the diagonal matrix ${\bm E}$ with elements $e_s$, we have $\bV^\dg
  {\bm C} \bV = {\bm E}~$. Comparing with Eq.~(\ref{unitary}) we see
  that the same
unitary transformation which diagonalizes ${\bm H}$ also diagonalizes the
correlation matrix ${\bm C}$ and we have ${\bm U}= {\bm V},~ {\bm
  D}={\bm E}$.

Using the results in Eqs.~(\ref{fermion_ss3}, \ref{fermion_ss4}) we then find,
$a_s=\ln (e_s^{-1}-1)$, and ${\bm A}= \bV^* ~{\rm Diag}~(a_1, a_2,\ldots,a_N)~
\bV^T$, which in turn  yields the steady state density matrix in
Eq.~(\ref{fermion_ss3}). For the equilibrium case $\mu_L=\mu_R=\mu,~T_L=T_R=T$
we have $d_s=e_s=f(\blam_s,\mu,T)~$, hence $a_s=(\blam_s -\mu)/ (k_BT) $ and ${\bm A}=[{\bm H}-\mu {\bm
    I}]/ (k_B T)$,  as expected. 
Thus we obtain, the non-trivial result, that the density matrix of a system, weakly coupled to two reservoirs at the same temperatures and chemical potentials, is given by the grand-canonical distribution of the isolated system. Note that this is not the case for the case of strong coupling.  

\subsection{Phonon transport}

\label{wc-phonon}
For the harmonic model we first note that there exists a
 real normal mode transformation matrix ${\bm V}$, with  elements ${\bm
   V}_l(s)$which satisfies:
\beq
{\bm V}^T {\bm M}{\bm V} = {\bm 1}~,~~~~{\bm V}^T {\bm K}{\bm V} = {\bm
  \Omega^2}~, \nn
\eeq
where ${\bm \Omega}$ is the diagonal matrix with elements as normal mode
frequencies.   It is easily verified that the matrix
\beq
{\bm S} &=&
\left(
\begin{array}{cc}
{\bm 0} &   -{\bm \Omega}^{-1/2} {\bm V}^T \\
{\bm \Omega}^{1/2} {\bm V}^T {\bm M}   & {\bm 0}
\end{array}
\right) ~\label{Seq}
\eeq
is symplectic {\emph {i.e}} ${\bm S} {\bm S}^T = {\bm J}$ and further has
the following property:
\beq
{\bm S} \left(
\begin{array}{cc}
{\bm K}^{-1} & {\bm 0}   \\
 {\bm 0} & {\bm M}
\end{array}
\right) ~{\bm S}^T =   \left(
\begin{array}{cc}
{\bm \Omega}^{-1} & {\bm 0}   \\
 {\bm 0} & {\bm \Omega}^{-1}
\end{array} \label{Seqprop}
\right)~.
\eeq

We now show that  the correlations for the harmonic system in
the weak coupling limit are given by:
\beq
\langle X~ X^T \rangle &=&   {\bm V}~ {\bm \Omega}^{-1/2}~{\bm E} ~{\bm
  \Omega}^{-1/2} ~ {\bm V}^T \label{xxneq} \\
\langle X~ P^T + [P~X^T]^T \rangle &=& 0 \label{xpneq} \\
\langle P~ P^T \rangle &=&
  {\bm M ~\bm V  }~{\bm \Omega}^{1/2} ~{\bm E}~
{\bm  \Omega}^{1/2}~ {\bm V}^T~{\bm M},  \label{ppneq}
\eeq
where we have defined the diagonal  matrix ${\bm E}$ whose elements
are given by:
\begin{eqnarray}
e_s  &=& \frac{\hbar}{2}~\sum_{a=L,R} \f{\langle
  ~s~|~\bGam_a~|~s~\rangle }{\langle   ~s~|~\bGam~|~s~\rangle }~
\coth \left[\f{\hbar  \bOm_s}{2 k_B
    T_a}\right]~~~~s=1,2,\ldots,N~ \label{wcphononD1}  \\
&=& \gamma_L ~\frac{\hbar}{2} ~ \coth \left[\f{\hbar  \bOm_s}{2 k_B     T_L}\right]+ \gamma_R ~\frac{\hbar}{2} ~ \coth \left[\f{\hbar  \bOm_s}{2 k_B     T_R}\right]~, \nn \\ 
{\rm where}~~ \gamma_{L} &=&  \f{\langle   ~s~|~\bGam_{L}~|~s~\rangle }{\langle   ~s~|~\bGam~|~s~\rangle }~,~~~~\gamma_{R} =  \f{\langle   ~s~|~\bGam_{R}~|~s~\rangle }{\langle   ~s~|~\bGam~|~s~\rangle }=1-\gamma_L~. \nn
\end{eqnarray}
Let us define the effective temperature $\widetilde{\bm T}_s$ for each
normal mode through the relation, reading:
\begin{eqnarray}
e_s &=& \frac{\hbar}{2}~\coth \left[ \frac{\hbar {\bm \Omega}_s}{2k_B \widetilde{\bm T}_s}\right]~, \label{wcphononD2} ~~~{\rm giving} \\
\frac{1}{\widetilde{\bm T}_s}&=& \frac{2k_B}{\hbar {\bm \Omega}_s}~\coth^{-1} \left[ \gamma_L ~ \coth \left(\f{\hbar  \bOm_s}{2 k_B     T_L}\right)+ \gamma_R  ~ \coth \left(\f{\hbar  \bOm_s}{2 k_B     T_R}\right) \right]~,  \nn 
\end{eqnarray}
which notably depends on both temperatures  $T_L$ and $T_R$. We remark here again that in the above expressions the limit $\langle ~
s~|~\bGam_a~|~s~\rangle \to 0 $ is
implied and it is then clear that the ratios  $\gamma_L$, $\gamma_R$ depend on
the way the couplings $\to 0$.

To prove the above results,
Eqs.~(\ref{xxneq}, \ref{xpneq}, \ref{ppneq}, \ref{wcphononD1}), we first
  introduce the following spectral decomposition:
\beq
\bG^+(\omega)&=&\bV \bV^{-1} [-\bM \omega^2 +\bK -\bSig_L^+-\bSig_R^+]^{-1}
   [\bV^T]^{-1} \bV^T \nn \\
&=& \bV ~ [ \bV^T~(~-\bM \omega^2 +\bK -\bSig_L^+-\bSig_R^+~)~\bV~]^{-1}~
   \bV^T \nn \\
&=& \bV ~ [ ~- \omega^2 +{\bm \Omega}^2  - \bV^T \bSig_L^+\bV - \bV^T
     \bSig_R^+\bV ~]^{-1}~    \bV^T \nn ~.
\eeq
From this it follows that in the weak coupling limit $\bSig_L^+,\bSig_R^+ \to
0$, the matrix element $\bG^+_{l,m}$ is effectively given by
\beq
\bG^+_{l,m}= \sum_{s} \f{\bV_l(s) \bV_m(s)}{-\omega^2 +\bOm_s^2 -i \langle
  ~s~|~\bGam~|~s~\rangle  }~,
\eeq
where $\langle   ~s~|~\bGam~|~s'~\rangle= \sum_{l,m} \bV_l(s)~ \bGam_{l,m}~ \bV_m(s')$
and  $\bGam=\bGam_L+\bGam_R$.
 It can be shown that the off-diagonal terms
$\langle~ s~|~\bGam ~|~s'~\rangle$ for $s \neq s'$, as well as the real part of $\bSig^+_{L,R}$
give lower  order contributions in the weak coupling limit and can be dropped.
Hence we find:
\beq
\langle~ x_l~ x_m ~\rangle &=& \int_{-\infty}^\infty \f{\hbar}{2 \pi} \sum_{a=L,R}\sum_{j,k}
\bG^+_{l,k} ~[\bGam_a]_{k,j}~ \bG^-_{j,m}~g(\omega,T_a)
 \nn \\
&=& \int_{-\infty}^\infty \f{\hbar}{2 \pi} \sum_{a=L,R}
\sum_{s,s',j,k} \f{\bV_l(s) \bV_k(s)}{-\omega^2 +\bOm_s^2 -i \langle~
  s~|~\bGam~|~s~\rangle }  [\bGam_a]_{k,j} \nonumber \\
&& ~~~~~~~~~~~~~~~~~~~~~\times
\f{\bV_j(s') \bV_m(s')}{-\omega^2 + {\bm \Omega}_{s'}^2 +i \langle~
  s'~|~\bGam~|~s'~\rangle  }~g(\omega,T_a)~. \nn
 \eeq
A careful examination of the limit $\langle ~s~|~\bGam_a~|~s~\rangle \to 0$
shows that only the terms $s=s'$ in the above summation survive and we then
obtain:
\beq
\langle~ x_l~ x_m~ \rangle =  \int_{-\infty}^\infty \f{\hbar}{2 \pi} \sum_{a=L,R}
\sum_{s} \f{\bV_l(s)~ \langle ~s~|~ \bGam_a(\omega)~ |~s~\rangle~ \bV_m(s) }{(-\omega^2
  +{\bm \Omega}_s^2)^2 + \langle  ~ s~|~\bGam(\omega)~|~s~\rangle^2 }
~g(\omega,T_a)~. \nn
\eeq
Now we note the following identity:
\beq
\lim_{\epsilon \to 0} \f{\epsilon}{(x^2-a^2)^2+\epsilon^2} = \f{\pi}{2 a} [
  \delta(x-a)+\delta(x+a) ]~.
\eeq
Using this and the fact that $\bGam_a(\omega)$ and $g(\omega)$ are both odd
functions of $\omega$, one arrives at:
\beq
\langle~ x_l~ x_m~ \rangle = \sum_s  \f{\hbar}{2} \bV_l(s) \bV_m(s)~\sum_{a=L,R} \f{\langle~
  s~|~\bGam_a~|~s~\rangle }{\langle ~  s~|~\bGam~|~s~\rangle }
\f{g(\bOm_s,T_a)}{\bOm_s}~,
\eeq
which proves Eq.~(\ref{xxneq}). Similarly we can evaluate other correlations
and obtain Eqs.(\ref{xpneq},\ref{ppneq}).

From  the form of the correlations in
  Eqs.~(\ref{xxneq}, \ref{xpneq}, \ref{ppneq}) we deduce that the matrix ${\bm
  S}$ given in Eq.~(\ref{Seq}) provides the   required symplectic
  transformation  in Eq.~(\ref{symplectic2}) with
 ${\bm   D}={\bm E}$.
Therefore, using Eq.~(\ref{boson_ss4}) and the definition in Eq.~(\ref{wcphononD2}) we get $a_s=\bOm_s/(2k_B \widetilde{\bm T}_s)$.
Finally Eq.~(\ref{boson_ss2b}) gives ${\bm A}= {\bm S}^T {\bm
  \Omega} \tilde{\bm T}^{-1} {\bm S}/(2k_B)$ and then from
  Eq.~(\ref{boson_ss2a}) we obtain $\rho_S$. 
This density matrix corresponds to each of the normal modes of the harmonic system being in equilibrium at an effective temperature $\widetilde{\bm T}_s$.
For the equilibrium case $T_L=T_R=T$, we find, using Eq.~(\ref{Seqprop}),  $\varphi^T {\bm A} \varphi= {\cal H}_S/(k_B T)$. This result is expected but is non-trivial, and is valid only in the weak-coupling limit.

\section{Application to  generic setups}
\label{sec:examples}
\subsection{Electron transport in a one-dimensional wire}
\subsubsection{System with single site}
We consider the system plus reservoir to consist of a single site, such as e.g. realized with a single-level quantum dot,
that is connected to two one-dimensional reservoirs. The full Hamiltonian then reads:

\beq
{\cal H} &=& {\cal H}_S + {\cal H}_L + {\cal H}_R +{\cal H}_{LS}+{\cal
  H}_{RS}\, , \nn \\
{\rm where }~~{\cal H}_S &=&  \epsilon c_0^\dg c_0~,\nn \\
{\cal H}_L &=& - \sum_{\alpha=1}^\infty \, t [~c_{\alpha}^{\dagger}
c_{\alpha+1} +  c_{\alpha+1}^{\dagger} c_{\alpha}~]~, ~~~
{\cal H}_R=- \sum_{\alpha'=1}^{\infty} \, t [~c_{\alpha'}^{\dagger}
c_{\alpha'+1} +  c_{\alpha'+1}^{\dagger} c_{\alpha'}~]~, \nn \\
{\cal H}_{LS}&=& - t'_L [ ~c_{\alpha=1}^\dg c_0 +  c_0^\dg c_{\alpha=1}~] ~,~~~{\cal H}_{RS}=- t'_R [ ~c_{\alpha'=1}^\dg c_0 +  c_0^\dg c_{\alpha'=1}~] ~.
\label{ham_fer_ex}
\eeq
The self-energies can be expressed in terms of the Green functions
of the uncoupled reservoir Hamiltonian ${\bm g}^+_{L,R}$ and the coupling
elements $t'_{L,R}$. Defining $\omega=-2 t \cos q$, where $0 \leq q \leq\pi$, we find that for $ |\omega|
\leq   2 t $:
\beq
\Sigma_L^+(\omega) &=& -\frac{{t'}^2_L}{t}e^{i q}
~,~~~~\Sigma_R^+(\omega)= -\frac{{t'}^2_R}{t}e^{i q}~, \label{1pSE1} \\
\Gamma_L^+(\omega) &=&  \frac{{t'}^2_L}{t}\sin{q}~,
~~~~\Gamma_R^+(\omega)= \frac{{t'}^2_R}{t}\sin{q}~. \nn
\eeq
Hence the system's Green function emerges to read:
\beq
G^+(\omega)=\frac{1}{\hbar\omega-\epsilon-\Sigma_L^+(\omega)-\Sigma_R^+(\omega)}~.
\eeq
The correlation matrix element for the single-site problem is then readily obtained, reading
given by:
\beq
d=\langle c_0^\dg c_0 \rangle = \int_{-2 t}^{2t} d \omega
\frac{\hbar}{\pi} |G^+(\omega)|^2 ~ [~\Gamma_L(\omega)~ f(\omega,\mu_L,T_L)
  +\Gamma_R(\omega)~ f(\omega,\mu_R,T_R) ~]~.
\eeq
Consequently we find for the steady state nonequilibrium density matrix for this case the explicit result

\beq
\rho_S &=&\frac{\exp (-a {c_0}^\dg c_0)}{1+\exp (-a)}  \label{ferm1PSS} \\
{\rm where}~~a&=& \ln (d^{-1}-1)~. \nn
\eeq

\subsubsection{System composed of two sites}
We next consider a system where the reservoirs are identical to those in the
previous section, while the system Hamiltonian and system-bath couplings are
as follows:
\beq
\cH_S &=& \epsilon_1 c^\dg_1 c_1 + \epsilon_2 c^\dg_2 c_2 -t (c^\dg_1 c_2+c^\dg_2
c_1) \nn \\
{\cal H}_{LS}&=& - t'_L [ ~c_{\alpha=1}^\dg c_1 +  c_1^\dg c_{\alpha=1}~] ~,~~~{\cal H}_{RS}=- t'_R [ ~c_{\alpha'=1}^\dg c_2 +  c_2^\dg c_{\alpha'=1}~] ~.
\eeq
The self energies are again given by Eq.~(\ref{1pSE1}) and the
system's Green function is then
\beq
\bG^+(\omega)= \left(
\begin{array}{cc}
\hbar\omega-\epsilon_1-\Sigma^+_L(\omega) &   t   \\
t & \hbar\omega-\epsilon_2-\Sigma^+_R(\omega)
\end{array}
\right)^{-1}~. \label{el2pG}
\eeq

In this case it is difficult to construct explicitly  the required unitary matrix ${\bm
  U}$  though it is straight-forward to evaluate it
numerically and from that find the steady state density matrix given by
Eq.~(\ref{fermion_ss1}).

In what follows we present  numerical precise results for this setup. In our numerics
we use the following set of parameter values: $t=1.0, t'_L=t'_R=0.05,
\epsilon_1=0.2, \epsilon_2=0.4, T_L=0.25, T_R=0.25$. The right
reservoir chemical potential is fixed at $\mu_R=0.0$ and we study the
NESS for different values of $\Delta \mu=\mu_L-\mu_R$.

The Green function in Eq.~(\ref{el2pG}) is first obtained and then
all the elements of the  correlation matrix  given by Eqs.~(\ref{negfres})
are evaluated by numerical integration. As examples we give below the
correlation matrices for the equilibrium case $\Delta \mu
=0$ and for $\Delta \mu= 2.0$.
\beq
{\bm C}_S &=& \left(
\begin{array}{cc}
0.519 ~&~  0.465 \\
0.465  ~&~  0.427
\end{array}
\right) ~~~{{\rm{ for}} ~~\Delta \mu=0}~, \nn \\
{\bm C}_S &=& \left(
\begin{array}{cc}
0.726 ~&~ 0.271 + i 0.000473  \\
0.271 - i 0.000473 ~&~  0.672
\end{array}
\right) ~~~{{\rm{ for}} ~~\Delta \mu=2.0}~. \nn
\eeq
The electron current in the chain is given by $j_e=2 t Im[\langle c_1^{\dg} c_2
\rangle] $ and in the above example $j_e=0.000946$.

As discussed  in Sec.(\ref{sec:ssel}) the NESS density matrix assumes the form:
\beq
\rho_S=\frac{\exp ({-c^\dg~ {\bm A}~ c})}{[1+\exp(-a_1)]~ [1+\exp(- a_2)]}~,
\eeq
where $c=(c_1,c_2)^T$ and we  numerically
determined the coefficients $a_1,a_2$ and the matrix ${\bm A}$.
Finding the eigenvalues and eigenvectors of ${\bm C}$ yields the
matrix ${\bm D}$ and the unitary  matrix ${\bm U}$, respectively.
We  evaluate  $a_1=\ln (d_1^{-1}-1), ~a_2=\ln  (
  d_2^{-1} -1 )$ and numerically  obtain  the steady state matrix
$${\bm A} = {\bm U}^\star ~Diag~(a_1,a_2)~{\bm U}^T.$$

Note that for $\Delta \mu=0$ ($\mu_L=\mu_R=0$) and with a  weak-coupling to
reservoirs, we expect the result,
$\rho_S =\rho_{eq} \sim e^{-\beta ( {\cal H}_S-\mu {\cal N}) }$ and hence
\beq
{\bm A}_{eq} &=& \left(
\begin{array}{cc}
 0.8 ~&~ -4.0  \\
 -4.0 ~&~ 1.6
\end{array}
\right)~.  \nn
\eeq
\begin{figure}
\includegraphics[width=14cm]{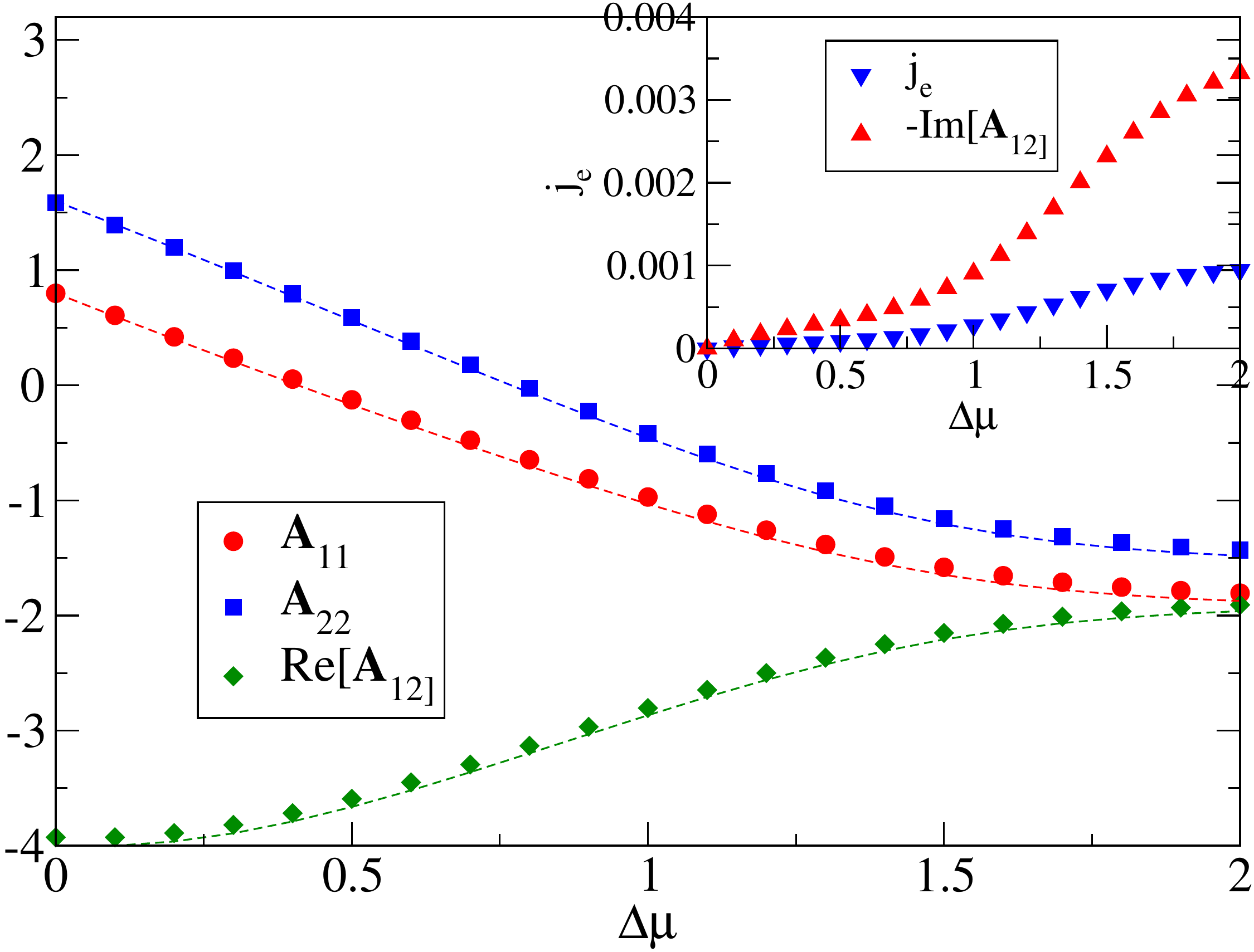}
\caption{
(color online). Plot of the  NESS matrix elements  ${\bm A}$ as a
  function of the chemical potential difference $\Delta \mu=\mu_L-\mu_R$ with
  fixed $\mu_R=0.0$ and with the remaining  parameters as given in the text. The dashed
  lines depict results obtained from the weak-coupling
  approximation. The inset shows the  electron current $j_e=2 Im[{\bm
      C}_{12}]$ together with $ Im[{\bm A}_{12}]$.}
\label{fig1}
\end{figure}
In Fig.~(\ref{fig1}) we depict the matrix elements
${\bm A}_{11},{\bm A}_{22}$ and $ Re[{\bm A}_{12}]$
as functions of the
chemical potential difference $\Delta \mu$.
In the inset we also evaluated
 the electron current; i.e.  $j_e=2~Im[{\bm
  C}_{12}] $ and show as well $ Im[{\bm A}_{12}]$ .

The  matrix elements
 ${\bm  A}_{11},{\bm A}_{22}$ and the real part of ${\bm A}_{12}$
can be obtained from our analytical weak-coupling results in
 Sec.~(\ref{wc-electron}). First
we obtain the  eigenvalues ${\bm \lambda}_s$ and
eigenfunctions ${\bm V}_l(s)$,$s=1,2$, corresponding to the isolated system
Hamiltonian ${\cal H}_S$. This provides the required
unitary transformation which diagonalises the matrix ${\bm C}$. For the present two-site setup the corresponding
eigenvalues, which determine the
matrix elements of ${\bm D}$,  generally given by
 Eqs.~(\ref{wcelectron}), take on  the following form:
\beq
d_s =  \frac{{t_L'}^2 |{\bm V}_1(s)|^2}{{t_L'}^2 |{\bm V}_1(s)|^2
 +{t'_R}^2
|{\bm V}_2(s)|^2 }~ \frac{1}{e^{(\lambda_s-\mu_L)/T_L}+1} +
 \frac{{t_R'}^2 |{\bm V}_2(s)|^2}{{t_L'}^2 |{\bm V}_1(s)|^2 +{t'_R}^2 |{\bm
              V}_2(s)|^2 }~\frac{1}{e^{(\lambda_s-\mu_R)/T_R}+1} \nn
\eeq
for $s=1,2$. The weak-coupling results for ${\bm A}_{11},{\bm A}_{22}$
and $ Re[{\bm A}_{12}]$  are depicted in Fig.~(\ref{fig1})  with dashed
lines. We notice that these are is excellent agreement with the  values obtained from exact
numerics.

\subsection{Phonon transport in one-dimensional oscillator chain}
\subsubsection{System consisting of a single oscillator}

We consider our system plus reservoir to be described by the full Hamiltonian
\beq
\cal{H} &=& \frac{p^2}{2 M} + \frac{k_o x^2}{2}  \nn \\
&+& \sum_{\alpha=1}^N \frac{p_\alpha^2}{2m} + \frac{k (x_\alpha-x_{\alpha+1})^2}{2} +
\frac{k'_L (x_{\alpha=1}-x)^2}{2} \nn \\
&+&  \sum_{\alpha'=1}^N \frac{p_{\alpha'}^2}{2m} + \frac{k (x_{\alpha'}-x_{\alpha'+1})^2}{2} +
\frac{k'_R (x_{\alpha'=1}-x)^2}{2} \nn~,
\eeq
where we assume $x_{\alpha=N+1}=x_{\alpha'=N+1}=0$. The above Hamiltonian can be written in
the canonical form:
\beq
\mathcal{H}&=&\mathcal{H}_S+\mathcal{H}_L+\mathcal{H}_R+\mathcal{H}_{LS}+\mathcal{H}_{RS}~,  \\
{\rm where}~~ \mathcal{H}_S &=&  \frac{p^2}{2 M} + \frac{(k_o+k'_L+k'_R) x^2}{2}~, \nn \\
\mathcal{H}_L&=& \sum_{\alpha=1}^N \frac{p_\alpha^2}{2m} + \frac{k
  (x_\alpha-x_{\alpha+1})^2}{2} +\frac{k'_L x_{\alpha=1}^2}{2}~, \nn \\
\mathcal{H}_R &=&  \sum_{\alpha'=1}^N \frac{p_{\alpha'}^2}{2m} + \frac{k (x_{\alpha'}-x_{\alpha'+1})^2}{2} +
\frac{k'_R x_{\alpha'=1}^2}{2} \nn~, \nn \\
\mathcal{H}_{LS}&=& -k'_Lx_{\alpha=1} x,~~~~~\mathcal{H}_{RS}= -k'_R x_{\alpha'=1}x ~.
\eeq

The self-energies can be expressed in terms of the Green functions of the
uncoupled reservoir Hamiltonian ${\bf g}^+_{L,R}(\omega)$ and the coupling
elements $k'_{L,R}$.
We define $\omega^2=(2k/m)~(1-\cos q)$, where $0\leq q \leq \pi$. Then, we find
that for $ |\omega| < \omega_m=  2(k/m)^{1/2}$:
\beq
\Sigma^+_L(\omega)&=&  \f{{k'_L}^2}{k} \f{\cos q - (1-u_L) + i \sin q}
      {2 (1-u_L) (1-\cos   q) +u_L^2 }~,~~  \Sigma^+_R(\omega)=
      \f{{k'_R}^2}{k} \f{\cos q - (1-u_R) + i \sin q} {2 (1-u_R) (1-\cos
  q) +u_R^2 }~,\nonumber \\
\Gamma_L(\omega) &=& \f{{k'_L}^2}{k} \f{ \sin q} {2 (1-u_L) (1-\cos
  q) +u_L^2 }~,~~\Gamma_R(\omega) =\f{{k'_R}^2}{k} \f{ \sin q} {2
  (1-u_R) (1-\cos  q) +u_R^2 }~,  \nn \\
&& \label{ph1pSE1}
\eeq
where $u_L = k'_L/k$ and $u_R=k'_R/k$. Hence the Green function is given by:
\beq
G^+(\omega)=\f{1}{-M \omega^2 +k_o+k_L'+k_R'-\Sigma^+_L(\omega)-\Sigma^+_R(\omega)}~.
\eeq
It is not difficult to verify that $\mathcal{T}(\omega)=4\Gamma_L(\omega)
\Gamma_R(\omega) |G^+(\omega)|^2 $ gives the correct transmission coefficient as can be
independently obtained by evaluating the transmission of plane waves from the left
reservoir to the right one, across the intermediate system.\\

The correlation matrix elements for the single-particle problem are obtained as:
\beq
c_1=\langle x^2 \rangle &=& \int_0^{\omega_m} d \omega \frac{\hbar}{\pi}
~|G^+(\omega)|^2 ~[~\Gamma_L(\omega)~g(\omega,T_L)+ \Gamma_R(\omega)~g(\omega,T_R)~]~, \nn \\
c_2=\langle p^2 \rangle &=& \int_0^{\omega_m} d \omega \frac{\hbar M^2 \omega^2}{\pi}~
|G^+(\omega)|^2~ [~\Gamma_L(\omega)~g(\omega,T_L)+ \Gamma_R(\omega)~g(\omega,T_R)~]~, \nn \\
\langle  x p+px  \rangle &=& 0~, \nn
\eeq
where $\omega_m=2(k/m)^{1/2}$ and $g(\omega,T)=\coth (\beta \hbar \omega/2)$.
Using the prescription in Sec.~(\ref{sec:ssph}) we find that $d_1=(c_1 c_2)^{1/2}$ and

\beq
{ \bm S} = \left(
\begin{array}{rr}
 0 & -(c_1/c_2)^{1/4} \\
(c_2/c_1)^{1/4} & ~ 0  \\
\end{array}
\right)~,
\eeq
yielding the explicit NESS density matrix:

\beq
\rho_S&=&\frac{e^{-[{\bm A}_{11}  x^2 + {\bm A}_{22} p^2 ]}}{Z} \nn \\
{\rm where}~~ {\bm A}_{11} &=& \left(\frac{c_2}{c_1}\right)^{1/2} ~a~,~~~{\bm
  A}_{22}~=~ \left(\frac{c_1}{c_2}\right)^{1/2} ~a~, \nn \\
a &=&\hbar^{-1} \coth^{-1} [ 2 \hbar^{-1} (c_1 c_2 )^{1/2}]~, \nn \\
Z &=&[2 \sinh (\hbar a) ]^{-1}~. \nn
\eeq

\subsubsection{System composed of two coupled oscillators}

In this case the baths have the same Hamiltonians as in the previous section
while the system Hamiltonian and system-bath couplings are given
by:
\beq
\mathcal{H}_S &=&  \frac{p_1^2}{2 m_1} + \frac{p_2^2}{2 m_2} +
 \frac{(k_1+k'_L) x_1^2}{2}  + \frac{k (x_1-x_2)^2}{2} + \frac{(k_2+k'_R)
    x_2^2}{2} \,, \nn \\
\mathcal{H}_{LS}&=& -k'_L x_{\alpha=1} x_1,~~~~~\mathcal{H}_{RS}= -k'_R
x_{\alpha'=1} x_2 ~.
\eeq
The self-energies are again given by Eq.~(\ref{ph1pSE1}) and the
system's Green function is
\beq
\bG^+(\omega)= \left(
\begin{array}{cc}
-m_1 \omega^2 + (k+k_1+k'_L) -\Sigma^+_L(\omega) &   -k   \\
-k & -m_2 \omega^2 + (k+k_2+k'_R) -\Sigma^+_R(\omega)
\end{array}
\right)^{-1}~. \label{ph2pG}
\eeq

For this setup it again becomes  difficult to evaluate explicitly  the symplectic matrix $ {\bm S}$ for
the general case though it is straight-forward to evaluate it numerically to yield
the steady state density matrix given by Eq.~(\ref{boson_ss2a}).

We  present some numerical results for this case. In our numerics we
fix the following parameter values: $m_1=1.0, m_2=1.5, k=k_1=k_2=1.0,
k_L'=k'_R=0.1$. Moreover, we keep the  temperature of the right reservoir fixed at
$T_R=1.0$ and study the NESS for different values of $\Delta T=T_L-T_R$.
We work in dimensionless units where $\hbar=k_B=1$. The temperatures $T_L,T_R$ are of the
order of the normal mode frequencies meaning  indeed that the system operates in the
quantum-mechanical regime.

The Green function in Eq.~(\ref{ph2pG}) is first obtained and then
all the elements of the  correlation matrix  given by Eqs.~(\ref{harmCM})
are evaluated by numerical integration. As examples we detail  below the
symmetric parts of the correlation matrices for the equilibrium case $\Delta
T=0$ and for $\Delta T= 4.0$.
\beq
{\bm C}_S &=& \left(
\begin{array}{cccc}
 0.696 ~&~ 0.294 ~&~ 0 ~&~ 0 \\
 0.294 ~&~ 0.670~&~ 0 ~&~ 0 \\
 0 ~&~ 0 ~&~ 1.168 ~&~ -0.0788 \\
 0 ~&~ 0 ~&~ -0.0788 ~&~ 1.67
\end{array}
\right) ~~~{{\rm{ for}} ~~\Delta T=0}~, \nn \\
{\bm C}_S &=& \left(
\begin{array}{cccc}
1.851 ~&~ 1.331 ~&~ 0 ~&~ -0.0294 \\
 1.331~&~ 2.241 ~&~ 0.0196 ~&~0 \\
 0 ~&~ 0.0196 ~&~ 2.491 ~&~ 0.781 \\
 -0.0294 ~&~0~&~ 0.781 ~&~ 4.558
\end{array}
\right) ~~~{{\rm{ for}} ~~\Delta T=4}~. \nn
\eeq
Note that the heat current across the chain is given by $j=k \langle x_1 p_2
\rangle/m_2 =-k \langle x_2 p_1 \rangle/m_1 = (k/m_2){\bm C}_{14} =
-(k/m_1){\bm C}_{23}$. For the above example  we obtain $j=0.0196$.

As shown in Sec.(\ref{sec:ssph}) the NESS density matrix assumes the form:
\beq
\rho_S=\frac{\exp ( -\varphi^T {\bm A} \varphi)}{4 \sinh (a_1) \sinh (a_2)}~,
\eeq
where $\varphi^T=(x_1,x_2,p_1,p_2)$. We  next numerically
determine $a_1,a_2$ and the matrix ${\bm A}$.
To this end we need to construct the diagonal matrix ${\bm D}$ and the symplectic
matrix ${\bm S}$. The way of constructing these are
described in Sec.~(\ref{appa}): It requires the following four  numerical procedures:

(i) Find the eigenvalues and eigenfunctions of ${\bm C}_S$.  Then
construct the matrix ${\bm C}_S^{1/2}$.

(ii) Find the eigenvalues and eigenvectors of the matrix $i {\bm
  C}_S^{1/2}{\bm J} {\bm C}_S^{1/2}$. There are four eigenvectors
  which occur as complex conjugate pairs,
  $(\omega_1^{+},~\omega_1^{-},~\omega_2^+,~\omega_2^-)$, with
  corresponding eigenvalues $(-d_1,~d_1,~-d_2,~d_2)$.

(iii) Evaluate the vectors $v_1^{\pm}={\bm C}_S^{1/2}
  \omega_1^{\pm}~, v_2^{\pm}={\bm C}_S^{1/2} \omega_2^{\pm}$ and use
Eqs.~(\ref{reim},\ref{vdef})  to  obtain the matrix
${\bm {\mathcal  V} }$.
The required  symplectic transformation is then ${\bm  S}=({\bm
  J} {\bm {\mathcal V}})^T$.

(iv) We evaluate  $a_1=\coth^{-1} (2 d_1), ~a_2=\coth^{-1} (2
  d_2)$ and  the steady state matrix
$${\bm A} = {\bm S}^T ~Diag~(a_1,a_2,a_1,a_2)~{\bm S}.$$

Note that for $\Delta T=0$ ($T_L=T_R=1$) and for  weak-coupling with
reservoirs, we expect
$\rho_S =\rho_{eq} \sim e^{-\beta {\cal H}_S}$;  hence
\beq
{\bm A}_{eq} &=& \left(
\begin{array}{cccc}
 1 ~&~ 0.5 ~&~ 0 ~&~ 0 \\
 0.5 ~&~ 1~&~ 0 ~&~ 0 \\
 0 ~&~ 0 ~&~ 0.5 ~&~ 0 \\
 0 ~&~ 0 ~&~ 0 ~&~ 0.33..
\end{array}
\right)~.  \nn
\eeq
\begin{figure}
\includegraphics[width=14cm]{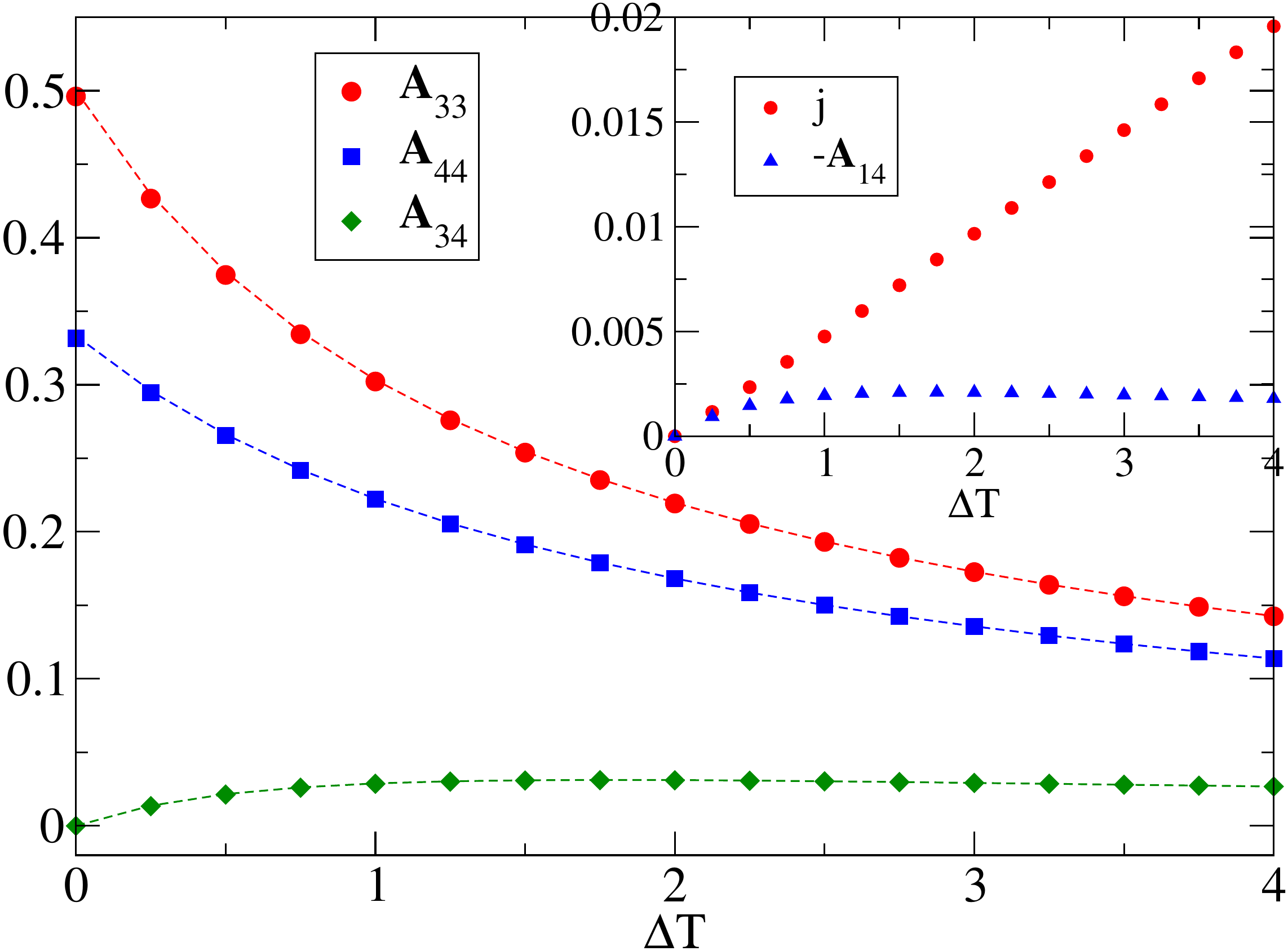}
\caption{
(color online). Plot of some relevant elements of the matrix ${\bm A}$ as a
  function of the temperature difference $\Delta T=T_L-T_R$ with
  constant $T_R=1.0$ while the other parameters are given in the text. The dashed
  lines depict results obtained from the analytical weak-coupling
  approximation. The inset shows both, the matrix element ${-\bm A}_{14}$ and the linearly growing
  heat current $j$.}
\label{fig2}
\end{figure}
In Fig.~(\ref{fig2}) we depict the matrix elements
${\bm A}_{33},{\bm A}_{44}$ and $ {\bm A}_{34}$
as functions of the
temperature difference $\Delta T$.
In the inset we have
plotted the element ${\bm A}_{14}$ and the heat current $j={\bm
  C}_{14}/m_2 $.\\

The $2 \times 2$ diagonal blocks of the matrix ${\bm A}$; i.e.,  ${\bm
  A}_{11},{\bm A}_{12},{\bm A}_{21}, {\bm A}_{22}$ and ${\bm A}_{33},{\bm A}_{34}, {\bm A}_{43},{\bm A}_{44}$,
can be obtained from the weak-coupling results in Sec.~(\ref{wc-phonon}). First
we obtain the normal mode eigenvalues ${\bm \Omega}_s$ and
eigenfunctions ${\bm V}_l(s)$,$s=1, 2$, corresponding to the isolated system
Hamiltonian ${\cal H}_S$. The
symplectic transformation is constructed by using Eq.~(\ref{Seq}). The
matrix elements of ${\bm D}$, given generally by Eqs.~(\ref{wcphononD1},\ref{wcphononD2}), takes the
following form:
\beq
d_s = \f{1}{2} \frac{{k_L'}^2 {\bm V}^2_1(s)}{{k_L'}^2{\bm V}^2_1(s) +{k'_R}^2 {\bm
              V}^2_2(s) }~\coth (\frac{\hbar {\bm \Omega}_s}{2k_BT_L}) +
\frac{1}{2} \frac{{k_R'}^2 {\bm V}^2_2(s)}{{k_L'}^2{\bm V}^2_1(s) +{k'_R}^2 {\bm
              V}^2_2(s) }~\coth (\frac{\hbar {\bm \Omega}_s}{2k_BT_R})~, \nn
\eeq
for $s=1, 2$. The weak-coupling results for ${\bm A}_{33},{\bm A}_{44}$
and $ {\bm A}_{34}$  have been plotted in Fig.~(\ref{fig2}) (dashed
lines) and we detect an excellent agreement with the  values obtained from precise
numerics.

\section{Conclusions and outlook}
\label{sec:discussions}

In summary,  we have detailed the explicit construction of the reduced density
matrix of the nonequilibrium steady states for two quantum transport problems,
one involving non-interacting fermionic degrees of freedom and the other noninteracting bosonic degrees.
The first setup concerns electron transport in a tight-binding lattice model
composed of non-interacting electrons that are connected to non-interacting baths while our
second setup focuses on heat transport across an arbitrary harmonic oscillator
network connected to harmonic oscillator baths.
For both these models the steady state correlations are known exactly
from various approaches and are usually expressed in terms of  nonequilibrium
Green  functions. We have demonstrated that for the Fermionic problem, the
construction of the emerging time-independent steady state density matrix requires that one evaluates a particular unitary
matrix while, likewise,  for the Bosonic case, it requires finding an appropriate
symplectic transformation.

For the  limiting case of vanishingly weak coupling between intermediate system and
reservoirs, we show that the required unitary and
symplectic transformations can be explicitly found and the resulting density matrices
assume simple forms whose explicit expressions depend on the way the coupling strengths are made to
vanish. For the case where the two baths possess the same temperatures (and
chemical potentials for electron case) the weak coupling case yields a unique
answer which is the expected equilibrium canonical (grand-canonical for
electrons) distribution. This requires the assumption that the connecting
reservoirs have sufficiently broad band-widths \cite{dharsen06,DharRoy06}.

The construction of the steady state density matrices required one to use
``diagonal'' representations [Eqs.~(\ref{fermion_ss1},\ref{boson_ss1})] and
these  are  analogous to  the eigenmode or
normal mode representation of the Hamiltonian.  In the equilibrium case and for weak coupling the
density matrix is $\sim e^{-\beta {\cal{H}}}$ and then the eigenmode
representation is useful in the computation of equilibrium averages of
various physical observables. Similarly, we expect that the ``diagonal''
representations  of the nonequilibrium density matrix  is as  useful for
computing averages in the NESS.
Thus, for example, the Von Neumann entropy of the nonequilibrium steady state,  defined as
$S = -{\rm Tr}~ [~\rho_{S} \ln \rho_{S}~]$ can be
readily obtained from our findings. In particular one finds that:\\
\begin{eqnarray}
\begin{array}{l}
S_{\rm fermion} = - \sum_{s=1}^N  (1-d_s ) \ln (1-d_s ) + d_s \ln d_s \, , \\
~\,S_{\rm boson} = -\sum_{s=1}^N
( {d_s/ \hbar} - {1/ 2} )
\ln ( {d_s / \hbar} - {1/ 2} )
-
( {d_s / \hbar} + {1/ 2} )
\ln ( {d_s / \hbar} + {1/ 2} ) \, ,
\end{array}
\end{eqnarray}
where $\{d_s\}$ are the ``diagonalized''  correlations defined via
Eqs.~(\ref{unitary}, \ref{symplectic2}).

\section*{Acknowledgments}
\noindent
We thank the Centre for Computational Science and Engineering,
National University of Singapore where this work was initiated.
AD thanks DST for support through the Swarnajayanti fellowship.
KS was supported by MEXT, Grant Number (23740289).
PH was supported by the DFG via SPP 1243  and via seed funding by
the excellence cluster ''Nanosystems Initiative Munich'' (NIM).

\appendix
\section{Procedure to find the symplectic matrix $\bm S$}
\label{appa}
We here explain the general procedure to find the symplectic matrix ${\bm S}$
\cite{gosson,gosson2}.
We first consider the eigenvalue problem for the matrices
$i{\bm C}_S^{1\over 2} {\bm J} {\bm C}_S^{1\over 2}$
and ${\bm C}_S {\bm J}$.
Note that the covariance matrix ${\bm C}_S$ is real-valued, symmetric and
 positive definite. Positive definitness is shown by ${y}^{T} {\bm C}_S {y}
={y}^{T} {\bm C} {y} = \langle ({\varphi}^{T} {y} )^2
\rangle_{ss}  \ge 0$ for arbitrary real column vector
${\bm y}$.

The matrix $i{\bm C}_S^{1\over 2} {\bm J} {\bm C}_S^{1\over 2}$ is
a Hermitian matrix. Therefore it possesses real eigenvalues as
$i{\bm C}_S^{1\over 2} {\bm J} {\bm C}_S^{1\over 2} {\omega}
= d {\omega} $,
where ${\omega}$ is the eigenvector. Taking the complex conjugate
of  both sides, we have the equation
$i{\bm C}_S^{1\over 2} {\bm J} {\bm C}_S^{1\over 2} {\omega}^{\ast}
= - d {\omega}^{\ast} $. From this, if $d$ is an eigenvalue, then  $- d$ is also an eigenvalue.

Hence, we can start with the following equations
\beq
i {\bm C}_S^{1\over 2} {\bm J} {\bm C}_S^{1\over 2}
\, {\omega}_k^{\pm} &=& \mp d_k {\omega}_k^{\pm}, \label{alg1}
\eeq
where ${\omega}_k^{\pm}$ are eigenvectors
$({\omega}_k^{-}={\omega}_k^{+}{}^{\ast})$
which have real eigenvalues $\mp d_k~\,(d_k > 0)$.
These equations are equivalent to
\beq
 {\bm C}_S {\bm J} {v}_k^{\pm}
&=& \pm i d_k  {v}_k^{\pm}  \label{s00}
\eeq
where the vectors ${v}_k^{\pm}$ are defined as
\beq
{v}_k^{\pm} &=& {\bm C}_S^{1\over 2}  \, {\omega}_k^{\pm} . \label{vkdef}
\eeq
We divide the vector ${v}_k^{\pm}$ into the real and imaginary parts as
\beq
{v}_k^{\pm} &=& {v}^R_k \pm i {v}^I_k . \label{reim}
\eeq
Then, Eq.(\ref{s00}) implies the two relations
\beq
{\bm C}_S {\bm J} {v}^R_k &=& - {v}^I_k  d_k  \, , \label{cbeta1}\\
{\bm C}_S {\bm J} {v}^I_k &=&   {v}^R_k  d_k .   \label{cbeta2}
\eeq
Because the matrix $i{\bm C}_S^{1\over 2} {\bm J} {\bm C}_S^{1\over 2}$
is Hermitian, we can normalize the vector ${\omega}_{k}^{\pm}$ as
\beq
({\omega}_{k}^{\pm})^{\dagger} {\omega}_{k '}^{\pm} &=& 2 d_{k'}^{-1}\delta_{k,k'} \, ,
\label{nm1} \\
({\omega}_{k}^{\pm})^{\dagger} {\omega}_{k '}^{\mp} &=& 0\, . \label{nm2}
\eeq
From (\ref{vkdef}), the vector $\omega_{k}^{\pm}$ is expressed with vectors
${v}_k^{R,I}$ as
\beq
\omega_{k}^{\pm} &=& {\bm C}_S^{-{1\over 2}}( {v}_k^{R} \pm i {v}_k^{I} ).
\eeq
Using this the Eqs.(\ref{nm1}) and (\ref{nm2}) are written as
\beq
( {v}_k^{R} \mp i {v}_k^{I} )^T {\bm C}_S^{-{1\over 2} }   {\bm C}_S^{-{1\over 2} }
( {v}_{k'}^{R} \pm i {v}_{k'}^{I} )
&=&
({v}_k^R )^T {\bm C}_S^{-1} {v}_{k'}^R +
({v}_k^I )^T {\bm C}_S^{-1} {v}_{k'}^I \nonumber \\
&\mp&
i \left[ ({v}_k^I )^T {\bm C}_S^{-1} {v}_{k'}^R -
({v}_k^R )^T {\bm C}_S^{-1} {v}_{k'}^I \right] = 2 d_{k'}^{-1} \delta_{k,k'}.  ~~~~~~  \\
( {v}_k^{R} \mp i {v}_k^{I} )^T {\bm C}_S^{-{1\over 2} }   {\bm C}_S^{-{1\over 2} }
( {v}_{k'}^{R} \mp i{v}_{k'}^{I} )
&=&
({v}_k^R )^T {\bm C}_S^{-1} {v}_{k'}^R -
({v}_k^I )^T {\bm C}_S^{-1} {v}_{k'}^I \nonumber \\
&\mp&
i \left[ ({v}_k^I )^T {\bm C}_S^{-1} {v}_{k'}^R +
({v}_k^R )^T {\bm C}_S^{-1} {v}_{k'}^I \right] = 0.
\eeq From this, we find the following  set of relations
\beq
({v}_k^R)^T {\bm C}_S^{-1} {v}_{k'}^R &=& d_{k'}^{-1} \delta_{k,k'} \, ,\\
({ v}_k^I)^T {\bm C}_S^{-1} { v}_{k'}^I &=& d_{k'}^{-1} \delta_{k,k'} \, ,\\
({ v}_k^R)^T {\bm C}_S^{-1} { v}_{k'}^I &=& 0 \, ,\\
({ v}_k^I)^T {\bm C}_S^{-1} { v}_{k'}^R &=& 0 \, .
\eeq
Utilizing Eqs.(\ref{cbeta1}) and (\ref{cbeta2}), the above relations can be recast as
\beq
({v}_k^R )^T {\bm J } {v}_{k'}^I &=& \delta_{k,k'} \, , \label{s01} \\
({v}_k^I )^T {\bm J } {v}_{k'}^R &=& -\delta_{k,k'} \, , \label{s02}\\
({v}_k^R )^T {\bm J } {v}_{k'}^R &=& 0 \, , \label{s03}\\
({v}_k^I )^T {\bm J } {v}_{k'}^I &=& 0 \, \label{s04}.
\eeq
We next define the $2N\times 2N$ matrix ${\bm{\mathcal{V}}}$
\beq
{{\bm {\mathcal V}}} &=& \left( {v}^R_1, \cdots, {v}^R_N , {v}^I_1, \cdots , {v}^I_N  \right) . \label{vdef}
\eeq
Using the matrix ${{\bm {\mathcal V}}}$, relations (\ref{cbeta1}) and (\ref{cbeta2})
can be simply written as
\beq
{\bm C}_S {\bm J}{{\bm {\mathcal V}}}
&=& {{\bm {\mathcal V}}}
{\bm J} {\bm D}
, ~~~\label{s1}
\eeq
where the matrix ${\bm D}$ is a $2N\times 2N$ diagonal matrix
\beq
{\bm D} &=& {\rm Diag} (d_1, \cdots , d_N , d_1, \cdots , d_N ) .
\eeq
In addition, the relations (\ref{s01})-(\ref{s04}) can be written with the matrix
${{\bm {\mathcal V}}}$ as
\beq
{{\bm {\mathcal V}}}^T {\bm J } {{\bm {\mathcal V}}} &=& {\bm J } . \label{s05}
\eeq
We now introduce the matrix ${\bm S}$ as
\beq
{\bm S} &=&
({\bm J}  {{\bm {\mathcal V}}} )^T. \label{sdef}
\eeq
One can prove that the matrix ${\bm S}$ satisfies the symplectic relation, namely:
\beq
{\bm S} {\bm J} {\bm S}^T &=&
 {{\bm {\mathcal V}}}^T {\bm J}^T
{\bm J } {\bm J }
 {{\bm {\mathcal V}}} \nonumber \\
&=&
-  {{\bm {\mathcal V}}}^{T}
 {\bm J}^T
 {{\bm {\mathcal V}}} = {{\bm {\mathcal V}}}^{T}
 {\bm J} {{\bm {\mathcal V}}} = {\bm J}, \label{c0} \\
{\bm S} {\bm C}_S {\bm S}^T &=&
{{\bm {\mathcal V}}}^T {\bm J}^T {\bm C}_S {\bm J} {{\bm {\mathcal V}}} \nonumber \\
&=&
{{\bm {\mathcal V}}}^T {\bm J}^T {{\bm {\mathcal V}}} {\bm J}  {\bm D}
 \nonumber \\
&=&
-{{\bm {\mathcal V}}}^T {\bm J} {{\bm {\mathcal V}}} {\bm J}  {\bm D}
 \nonumber \\
&=&
-{\bm J}^2 {\bm D}
= {\bm D}, \label{c4}
\eeq
where we used Eqs.(\ref{s1}) and (\ref{s05}).

To evaluate the symplectic matrix ${\bm S}$ numerically, we first solve
eigenvalue problem (\ref{alg1}) to obtain the eigenfunction ${\omega}_{k}^{\pm}$.
Next, we normalize them as in (\ref{nm1}), and find ${v}_k^{R,I}$.
Finally, constructing the matrix ${{\bm {\mathcal V}}}$ as in (\ref{vdef}), 0ne obtains the symplectic
matrix as in (\ref{sdef}).


\begin{thebibliography}{10}

\bibitem{PR} P.~H\"anggi and H. Thomas, {\it Stochastic Processes:
  Time-Evolution, Symmetries and Linear Response }, Phys. Rep. {\bf 88}, 207 (1982); cf. Sects. (1.3), (3.3),  (4.4) and (6.) therein.

\bibitem{CampisiPRL}
M. Campisi, P. Talkner, and P. H\"anggi, {\it Fluctuation theorem for arbitrary open quantum systems}, Phys. Rev. Lett. \textbf{102}, 210401 (2009); see Eq. (11) therein.

\bibitem{RLL67} Z.~Rieder, J.~L. Lebowitz, and E.~Lieb,
{\it Properties of a harmonic crystal in a stationary nonequilibrium
  state\/}, J. Math. Phys.~{\bf 8}, 1073 (1967).
\bibitem{Nakazawa} H. Nakazawa, {\it Energy Flow in Harmonic Linear
  Chain}, Progress of Theoretical Physics~ {\bf 39}, 236 (1968);
   {\it On the Lattice Thermal
  Conduction}, Progress of Theoretical Physics Supplement {\bf 45},
  231 (1970).
\bibitem{zurcher90} U. Z\"urcher and P. Talkner, {\it Quantum-mechanical
  harmonic chain attached to heat baths. II. Nonequilibrium
  properties}, Phys. Rev. A {\bf 42}, 3278 (1990).
\bibitem{saito00} K. Saito, S. Takesue, and S. Miyashita, {\it Energy
 transport in the integrable system in contact with various types of
 phonon reservoirs},  Phys. Rev. E {\bf 61}, 2397 (2000).
\bibitem{dharshastry03} A. Dhar and B. S. Shastry, {\it Quantum transport using the Ford-Kac-Mazur formalism}, Phys. Rev. B {\bf 67},
  195405 (2003).
\bibitem{segal03} D. Segal, A. Nitzan and P. H\"anggi, {\it Thermal
  conductance through molecular wires}, J. Chem. Phys. {\bf 119}, 6840 (2003).

\bibitem{zubarev74} D.N. Zubarev, {\em Nonequilibrium Statistical
  Thermodynamics}, (Consultants Bureau, New York, 1974).

\bibitem{mclennan63}  J.A. McLennan Jr., {\it The Formal Statistical Theory of
  Transport Processes}, Adv. Chem. Phys. {\bf 5}, 261 (1963).

\bibitem{pillet} V. Jaksic, Y. Ogata and C. A. Pillet, {\it Mathematical
  theory of non-equilibrium quantum statistical mechanics},
  J. Stat. Phys. {\bf 108}, 787 (2002).


 \bibitem{tasaki} S. Tasaki and J. Takahashi, {\it Nonequilibrium Steady
   States and MacLennan-Zubarev Ensembles in a Quantum Junction System},
Prog. Theor. Phys. Suppl. {\bf 165}, 57 (2006).

\bibitem{platini} D. Karevski and T. Platini   {\it Quantum Nonequilibrium
  Steady States Induced by Repeated  Interactions }, Phys. Rev. Lett. {\bf
  102}, 207207 (2009).

\bibitem{jauho94} A.-P. Jauho, N. S, Wingreen, and Y. Meir, {\it
  Time-dependent transport in interacting and noninteracting
  resonant-tunneling systems}, Phys. Rev. B {\bf 50}, 5528 (1994).

\bibitem{haugjauho} H. Haug and A.-P. Jauho, {\it Quantum Kinetics in
  Transport and Optics of Semiconductors} (Springer, Berlin, 1996).


\bibitem{dharsen06} A. Dhar and D. Sen, {\it Nonequilibrium Green's function
  formalism and the problem of   bound states},  Phys. Rev. B {\bf 73}, 085119
(2006).
\bibitem{DharRoy06} A. Dhar and D. Roy, {\it Heat transport in
  Harmonic lattices}, J.~Stat.~Phys.~{\bf 125}, 801 (2006).

\bibitem{Dhar08} A.~Dhar, { \it Heat Transport in low-dimensional systems}, Adv.~ Phys. {\bf 57}, 457 (2008).




\bibitem{tasaki01} S. Tasaki, {\em Nonequilibrium Stationary States of
  Nonintercting Electrons in a one-dimensional lattice}, Chaos, Solitons and
  Fractals {\bf 12}, 2657 (2001).

\bibitem{caroli} C. Caroli, R. Combescot, P. Nozieres, and D. Saint-James,
  {\it Direct calculation of the tunneling current},
J. Phys. C {\bf 4}, 916 (1971).

\bibitem{meir} Y. Meir and N. S. Wingreen, {\it Landauer formula for the
  current through an interacting electron region}, Phys. Rev. Lett. {\bf 68},
  2512 (1992).



\bibitem{wang06} J.-S. Wang, J. Wang and N. Zeng, {\it Nonequilibrium
  Green's function approach to mesoscopic thermal transport},
Phys. Rev. B 74, 033408 (2006); J.-S. Wang, J. Wang and J.T. Lu, {\it Quantum thermal transport in nanostructures}, Euro. Phys. Jn. B {\bf 62}, 381 (2008).
\bibitem{yamamoto06} T. Yamamoto and K. Watanabe, {\it Nonequilibrium
  Green's function approach to phonon transport in defective carbon
  nanotubes}, Phys. Rev. Lett. {\bf 96}, 255503 (2006).

\bibitem{gosson} M. de Gosson, {\it Symplectic geometry and quantum mechanics},
(Birkhauser Verlag, Berlin, 2006).
\bibitem{gosson2} M. de Gosson and F. Luef, {\it Symplectic capacities and the geometry of uncertainty: The irruption of symplectic topology in classical and quantum mechanics}, Phys. Rep. {\bf 484}, 131 (2009).
\end{thebibliography}
\end{document}